\documentclass[reprint,twocolumn,superscriptaddress,noshowpacs,notitlepage,longbibliography,10pt]{revtex4-2}%

\usepackage{graphicx,bm,times}
\usepackage{amsmath}
\usepackage{mathtools}
\usepackage{amsfonts}
\usepackage{amssymb}
\usepackage{color}
\usepackage{hyperref}
\DeclareUnicodeCharacter{2212}{-}
\hypersetup{
	colorlinks = true,
	allcolors = {blue}
}

\begin{document}

\title{Orbital-selective Band Hybridisation at the Charge Density Wave Transition in Monolayer TiTe$_2$}

\author{Tommaso Antonelli}
\affiliation{SUPA, School of Physics and Astronomy, University of St Andrews, St Andrews KY16 9SS, United Kingdom}

\author{Warda Rahim}
\affiliation{Department of Chemistry and Thomas Young Centre, University College London, 20 Gordon Street, London, WC1H 0AJ, United Kingdom}

\author{Matthew D. Watson}
\affiliation{SUPA, School of Physics and Astronomy, University of St Andrews, St Andrews KY16 9SS, United Kingdom}
\affiliation{Diamond Light Source, Harwell Campus, Didcot, OX11 0DE, United Kingdom}

\author{Akhil Rajan}
\author{Oliver J. Clark}
\author{Alisa Danilenko}
\altaffiliation[Present address: ]{Center for Quantum Devices, Niels Bohr Institute, University of Copenhagen, 2100 Copenhagen, Denmark}
\affiliation{SUPA, School of Physics and Astronomy, University of St Andrews, St Andrews KY16 9SS, United Kingdom}

\author{Kaycee Underwood}
\affiliation{SUPA, School of Physics and Astronomy, University of St Andrews, St Andrews KY16 9SS, United Kingdom}
\author{Igor Markovi{\'c}}
\author{Edgar Abarca-Morales}
\affiliation{SUPA, School of Physics and Astronomy, University of St Andrews, St Andrews KY16 9SS, United Kingdom}
\affiliation {Max Planck Institute for Chemical Physics of Solids, N{\"o}thnitzer Stra{\ss}e 40, 01187 Dresden, Germany}

\author{Se{\'a}n R. Kavanagh}
\affiliation{Department of Chemistry and Thomas Young Centre, University College London, 20 Gordon Street, London, WC1H 0AJ, United Kingdom}
\affiliation{Department of Materials and Thomas Young Centre, Imperial College London, Exhibition Road, London SW7}

\author{P. \surname{Le F\`{e}vre}}
\author{F. \surname{Bertran}}
\affiliation{Synchrotron SOLEIL, CNRS-CEA, L’Orme des Merisiers, Saint-Aubin-BP48, 91192 Gif-sur-Yvette, France}

\author{K. \surname{Rossnagel}}
\affiliation{Institut f{\"u}r Experimentelle und Angewandte Physik, Christian-Albrechts-Universit{\"a}t zu Kiel, 24098 Kiel, Germany}
\affiliation{Ruprecht Haensel Laboratory, Deutsches Elektronen-Synchrotron DESY, 22607 Hamburg, Germany}

\author{David~O.~Scanlon}
\affiliation{Department of Chemistry and Thomas Young Centre, University College London, 20 Gordon Street, London, WC1H 0AJ, United Kingdom}

\author{Phil~D.~C.~King}
\email{philip.king@st-andrews.ac.uk}
\affiliation{SUPA, School of Physics and Astronomy, University of St Andrews, St Andrews KY16 9SS, United Kingdom}

\begin{abstract}
An anomalous $(2\times2)$ charge density wave (CDW) phase emerges in monolayer 1T-TiTe$_2$ which is absent for the bulk compound, and whose origin is still poorly understood. Here, we investigate the electronic band structure evolution across the CDW transition using temperature-dependent angle-resolved photoemission spectroscopy. Our study reveals an orbital-selective band hybridisation between the backfolded conduction and valence bands occurring at the CDW phase transition, which in turn leads to a significant electronic energy gain, underpinning the CDW transition. For the bulk compound, we show how this energy gain is almost completely suppressed due to the three-dimensionality of the electronic band structure, including via a $k_z$-dependent band inversion which switches the orbital character of the valence states. Our study thus sheds new light on how control of the electronic dimensionalilty can be used to trigger the emergence of new collective states in 2D materials.

\end{abstract}
\date{\today}
\maketitle

\section*{Introduction}

\begin{figure*}
	\centering
	\includegraphics[width=\linewidth]{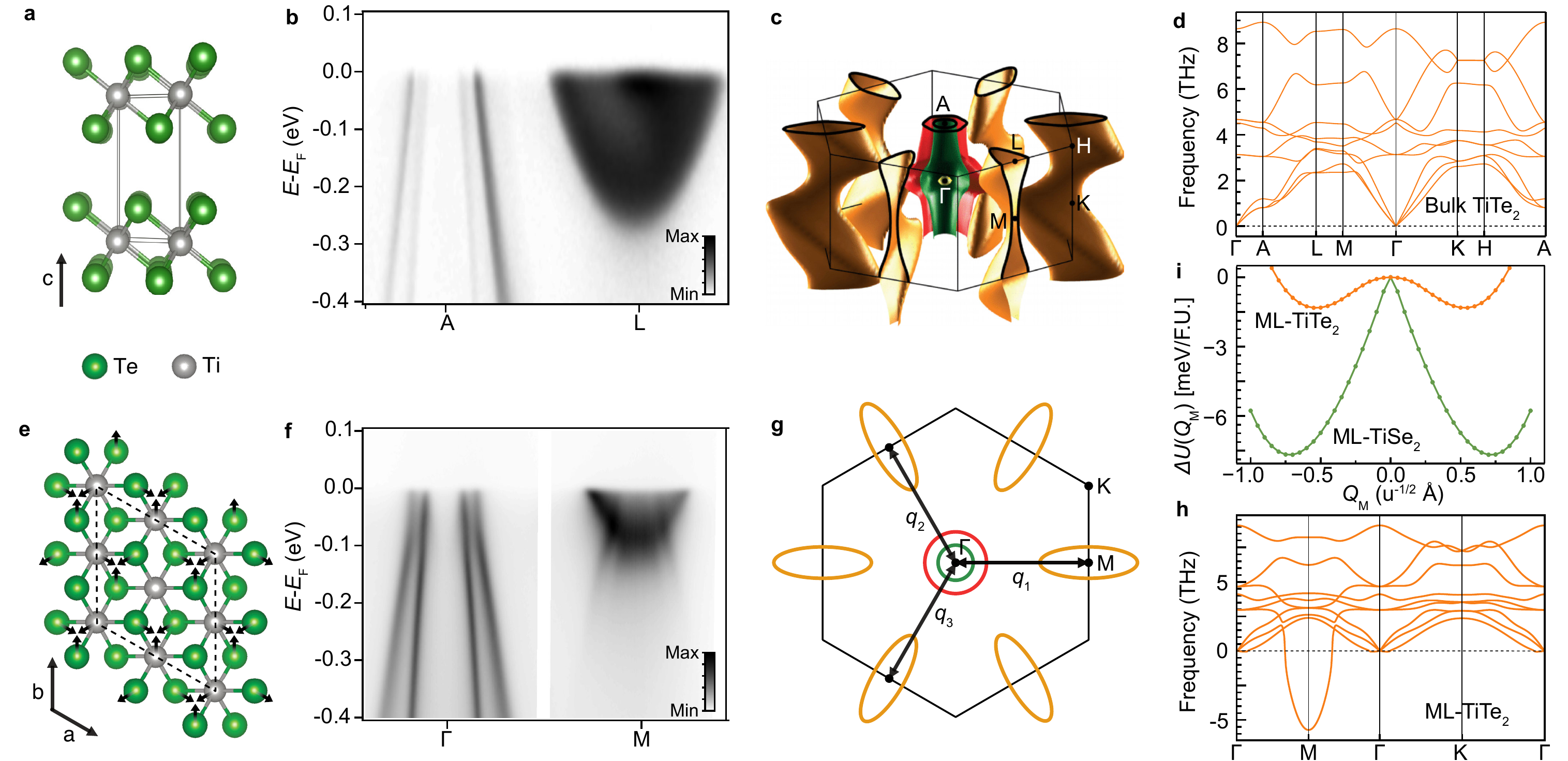}
	\caption{(a) Crystal structure of bulk 1T-TiTe$_2$ and (b) its measured electronic structure from ARPES ($h\nu = 144$~eV, $T=20$~K) along the A-L direction of the Brillouin zone. (c) Illustration of the bulk Fermi surface. (d) Calculated phonon dispersion of bulk TiTe$_2$, indicating stability of the 1T crystal structure. (e) Crystal structure of ML-TiTe$_2$, where the black arrows indicate the expected direction of the atomic displacements of the periodic lattice distortion associated with a CDW instability. (f) ARPES spectrum of ML-TiTe$_2$ measured along the $\Gamma$-M direction ($h\nu$ = 20 eV, $T=16$~K). (g) Schematic Fermi surface of ML-TiTe$_2$, with the three CDW wave vectors $q_1$, $q_2$ and $q_3$, indicated. (h) Calculated phonon dispersions of ML-TiTe$_2$, indicating the instability of the 1T crystal structure. (i) Tracking the potentials energy curve $\Delta U$ as a function of the corresponding normal-mode coordinates $Q_M$, with finite but small energy gain for the distortion; corresponding calculations for ML-TiSe$_2$ (green) indicate a much more pronounced instability at the same wavevector as for the Te case (orange).}
	\label{fig1}
\end{figure*}

Transition Metal Dichalcogenides (TMDs) offer a versatile platform to study the interplay of different collective quantum states \cite{Wang2012, Chhowalla2013, Bahramy2018}. Excitingly, the charge density wave (CDW) phases in these materials have been shown to exhibit a strong dependence on material thickness down to the monolayer (ML) limit, with the emergence of modified ordering wavevectors \cite{DuvjirNano18} and transition temperatures \cite{Xi2015a}, and evidence for significant phase competition~\cite{FengNanoLett}. A particularly intriguing case is that of TiTe$_2$. This is the sister compound of the famous charge-density wave material TiSe$_2$ \cite{DiSalvo1976PRB, Rossnagel2002}, in which an unconventional $(2\times2\times2)$ CDW emerges from a narrow-gap semiconducting normal state~\cite{Watson2019PRL}. CDW order persists down to the monolayer limit, with a $(2\times2)$ ordering~\cite{Chen2015_NatComms_monolayers}. In TiTe$_2$, a larger $p$-orbital valence bandwidth (due to the more extended Te 5$p$ orbitals) and enhanced $d$-$p$ charge transfer (promoted by the lower electronegativity of Te) cause the Ti $d$-derived conduction band and Te $p$-derived valence band to substantially overlap (Fig.~\ref{fig1}(a-c)). This semi-metallic state persists in the bulk to low temperatures, and no CDW transition occurs \cite{Koike1983, AllenPhysRevB, Claessen}.

Surprisingly, however, signatures of a CDW phase have recently been observed in monolayer TiTe$_2$, although not in bilayer or thicker films \cite{Chen2017}. This new CDW is associated with a $(2\times2)$ periodic lattice distortion, in striking similarity to the CDW observed in ML-TiSe$_2$ (Fig.~\ref{fig1}(e-g)), despite the significant band overlap and semimetallicity in the normal state of the Te-based system. The origins of this instability, and why it is only stabilised in the monolayer limit, have remained elusive to date. Here, we study the CDW transition in ML-TiTe$_2$ using temperature-dependent angle-resolved photoemission spectroscopy (ARPES), hybrid density functional theory (DFT) calculations, and minimal-model based approaches. Through this, we find evidence of a marked but symmetry-selective band hybridisation between the valence and conduction bands which ultimately provides the energetic incentive to drive the CDW phase transition. 
For the bulk system, we demonstrate how pronounced out-of-plane band dispersions lead to mismatched Fermi surfaces and inverted valence band symmetries. As a result, the corresponding electronic energy gain is highly suppressed, explaining the marked difference in the ground state properties of single vs.\ multi-layer TiTe$_2$.

\section*{Results}
\subsection*{CDW state in ML-T\lowercase{i}T\lowercase{e}$_2$}
To confirm the previously reported CDW state in ML-TiTe$_2$ as an intrinsic instability of the perfect ML, we start with DFT calculations of the lattice dynamics. The calculated phonon dispersions of bulk 1T-TiTe$_2$ are shown in Fig.~\ref{fig1}(d). These show an absence of any imaginary frequency modes: the 1T crystal structure thus is predicted to remain stable down to 0~K. This is consistent with the absence of any CDW transition in our measurements of the electronic structure from bulk TiTe$_2$ in Fig.~\ref{fig1}(b) and with previous experiments \cite{Koike1983,AllenPhysRevB, Claessen}. In contrast, for the monolayer (Fig.~\ref{fig1}(h)), our calculations indicate that the 1T crystal structure is no longer stable. The imaginary frequency mode at the M point indicates the tendency of the structure to undergo a $(2\times2)$ lattice distortion. This phonon mode has an $A_u$ irreducible representation at M, consistent with the soft phonon mode thought to underpin the triple-$q$ CDW instability in ML-TiSe$_2$~\cite{Kaneko2018PRB}. Tracing the stability of this mode (Fig.~\ref{fig1}(i)), we find a classic anharmonic double-well potential, confirming the intrinsic instability of ML-TiTe$_2$ to undergoing a periodic lattice distortion upon cooling. 

The underlying lattice instability is qualitatively the same as that in the sister compound ML-TiSe$_2$ (Fig.~\ref{fig1}(i)), but we find that the depth of the potential well in the telluride is much smaller. This indicates that the instability in ML-TiTe$_2$ is intrinsically weaker than that of TiSe$_2$, where signatures of a strong-coupling CDW transition have been observed~\cite{Watson_2020,Chen2015_NatComms_monolayers}. Nonetheless, consistent with a recent work~\cite{Chen2017}, we find that the new periodic potential arising from the softening of the M-point phonon mode in ML-TiTe$_2$ is sufficiently strong to induce a `backfolded' copy of the valence band dispersions, evident at the original M point in our measurements of the low-temperature electronic structure (Fig.~\ref{fig1}(f)). The backfolded spectral weight is weaker than the corresponding spectral signatures in ML-TiSe$_2$~\cite{Watson_2020}, qualitatively consistent with a `weaker' instability in the telluride as discussed above. Nonetheless, its observation points to a clear electronic component to the predicted periodic lattice distortion. 

\begin{figure*}
	\centering
	\includegraphics[width=\linewidth]{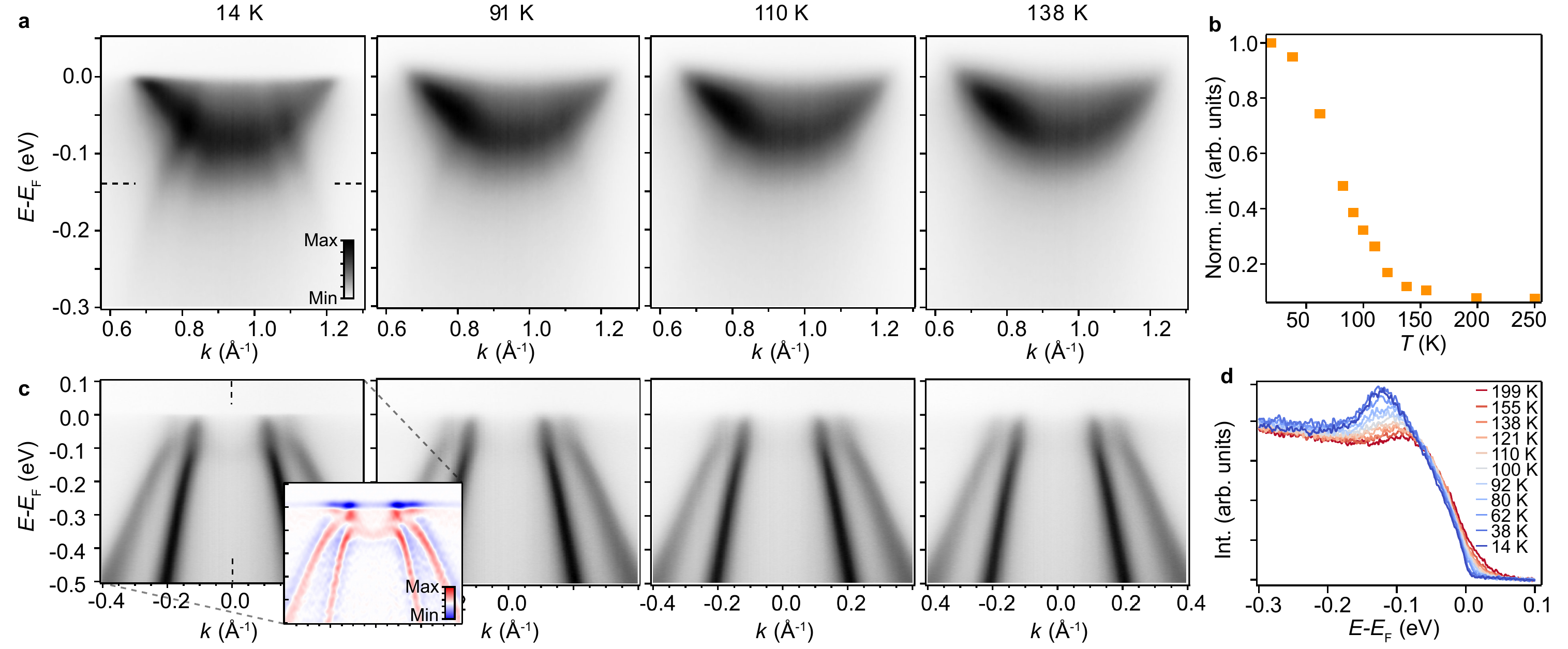}
	\caption{Temperature-dependent ARPES measurements of ML-TiTe$_2$ ($h\nu=15$~eV) along the $\Gamma$-M direction and centered at the (a) M and (c) $\Gamma$ points. (b) Intensity of the backfolded valence bands extracted from fits to MDCs at $E-E_\mathrm{F}=-0.14$~eV (along the dashed line in (a)). (d) temperature-dependent EDCs extracted at $k = 0$ (dashed line in (c)). The inset for the 14~K spectrum in (c) shows a second-derivative analysis of the measured band dispersion.}
	\label{fig2}
\end{figure*}

To investigate this in detail, we study the temperature-dependent evolution of the electronic structure of ML-TiTe$_2$ in Fig.~\ref{fig2}. With decreasing temperature, the spectral weight of the characteristic backfolded valence bands at the M-point gradually increases (Fig.~\ref{fig2}(a)). This is evident in the spectral weight extracted from fits to momentum distribution curves (MDCs) at $E-E_F$ = -0.14~eV (Fig.~\ref{fig2}(b), see Supplemental Fig.~S1 for the fitting details), where a rapid increase of the backfolded weight is evident at a temperature $T\approx110$~K. This is similar, although slightly higher, than the value of 92~K reported by Chen et al. \cite{Chen2017}, and we assign this as the CDW transition temperature in this system, $T_{\mathrm{CDW}}$. We note that this critical temperature is substantially lower than  $T_{\mathrm{CDW}}\approx220$~K in ML-TiSe$_2$~\cite{Watson_2020}. This is qualitatively consistent with the calculated double-well potentials shown in Fig.~\ref{fig1}(i): the depths of these wells are related to the thermal energy at which the distorted phase is no longer stable as compared to the high-temperature structure, located at the saddle point of these wells. The smaller depth of the potential well in the case of ML-TiTe$_2$ than TiSe$_2$, therefore, indicates that the CDW ordering would be expected to onset at a lower temperature in the former, as observed here.

The $(2\times2)$ order predicted by our calculations is in agreement with a superstructure modulation imaged by scanning tunnelling microscopy in Ref.~\cite{Chen2017}. This naturally explains the valence band backfolding from the $\Gamma$ to the M-point of the Brillouin zone, although a concomitant backfolding of the conduction band from M to $\Gamma$ has not been observed to date. Here, we find the appearance of weak spectral weight in between the innermost valence band in the ordered state (Fig.~\ref{fig2}(c)). Its spectral weight grows with deceasing temperature, as evident from a peak at $\approx100$~meV binding energy in energy distribution curves (EDCs) taken at the Brillouin zone centre (Fig.~\ref{fig2}(d)). A second-order derivative analysis (inset of Fig.~\ref{fig2}(c)) indicates how this additional spectral weight derives from an electron-like band, suggesting its origin as due to a backfolding of the conduction bands in the CDW phase of ML-TiTe$_2$.

\begin{figure}[!t]
	\includegraphics[width=\columnwidth]{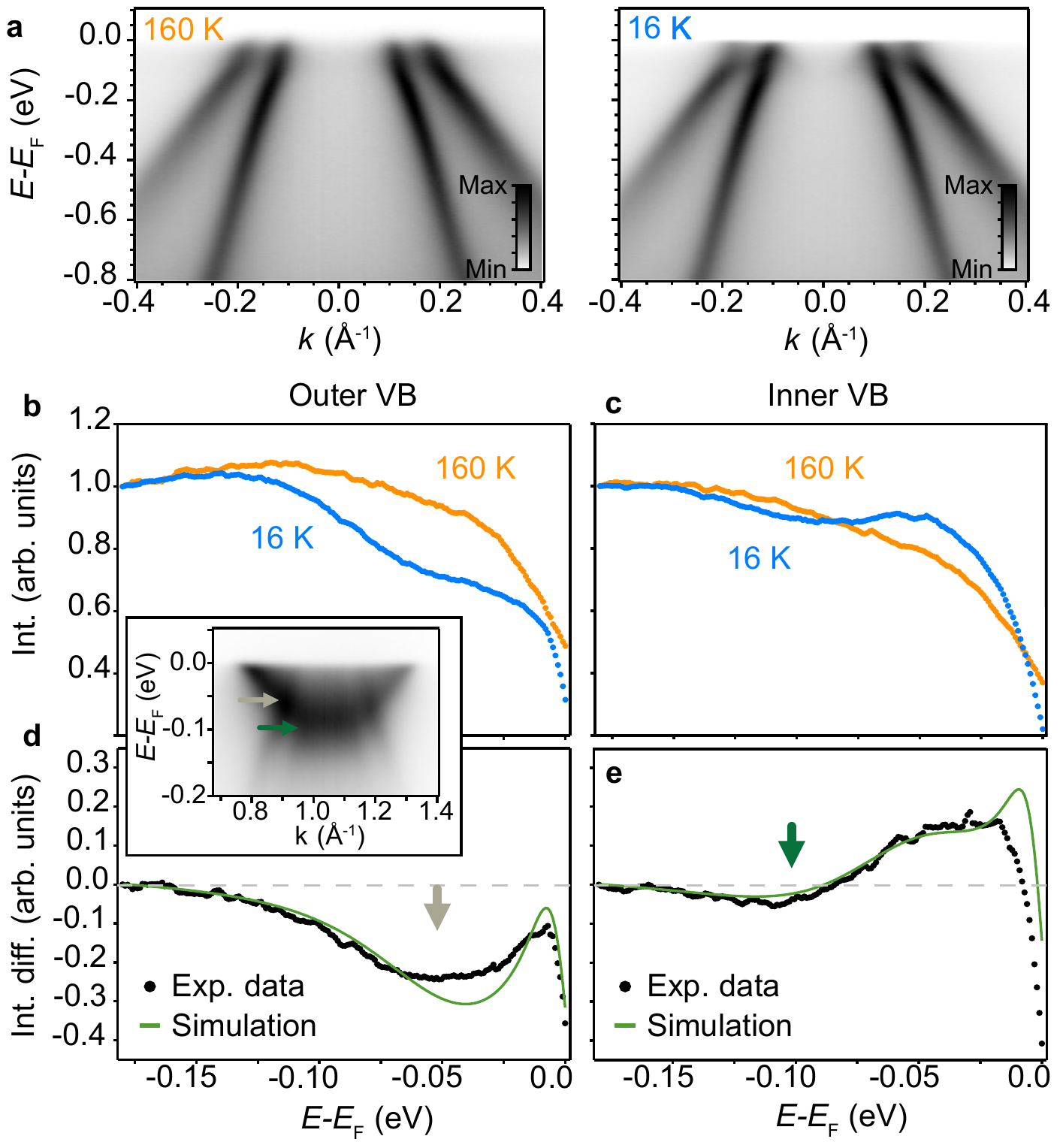}
	\caption{(a) ARPES spectra ($h\nu=20$~eV) measured around $\Gamma$ along the $\Gamma$-M direction in the normal (left, $T=160$~K) and CDW (right, $T=16$~K) state. (b,c) Intensities extracted from fits to MDCs in the vicinity of the Fermi level from the dispersions in (a), shown for the (b) inner and (c) outer valence bands. (d,e) Intensity difference between the MDC fits to the high and low temperature ARPES spectra (black dots). The green lines are the corresponding intensity difference extracted from the simulated spectra shown in Fig.~\ref{fig5}(c, e). The grey and green arrows indicate the energy where the backfolded conduction and valence bands intersect (shown in the inset from a measurement around the M point).}
	\label{fig3}
\end{figure}

\begin{figure}[!b]
	\includegraphics[width=\columnwidth]{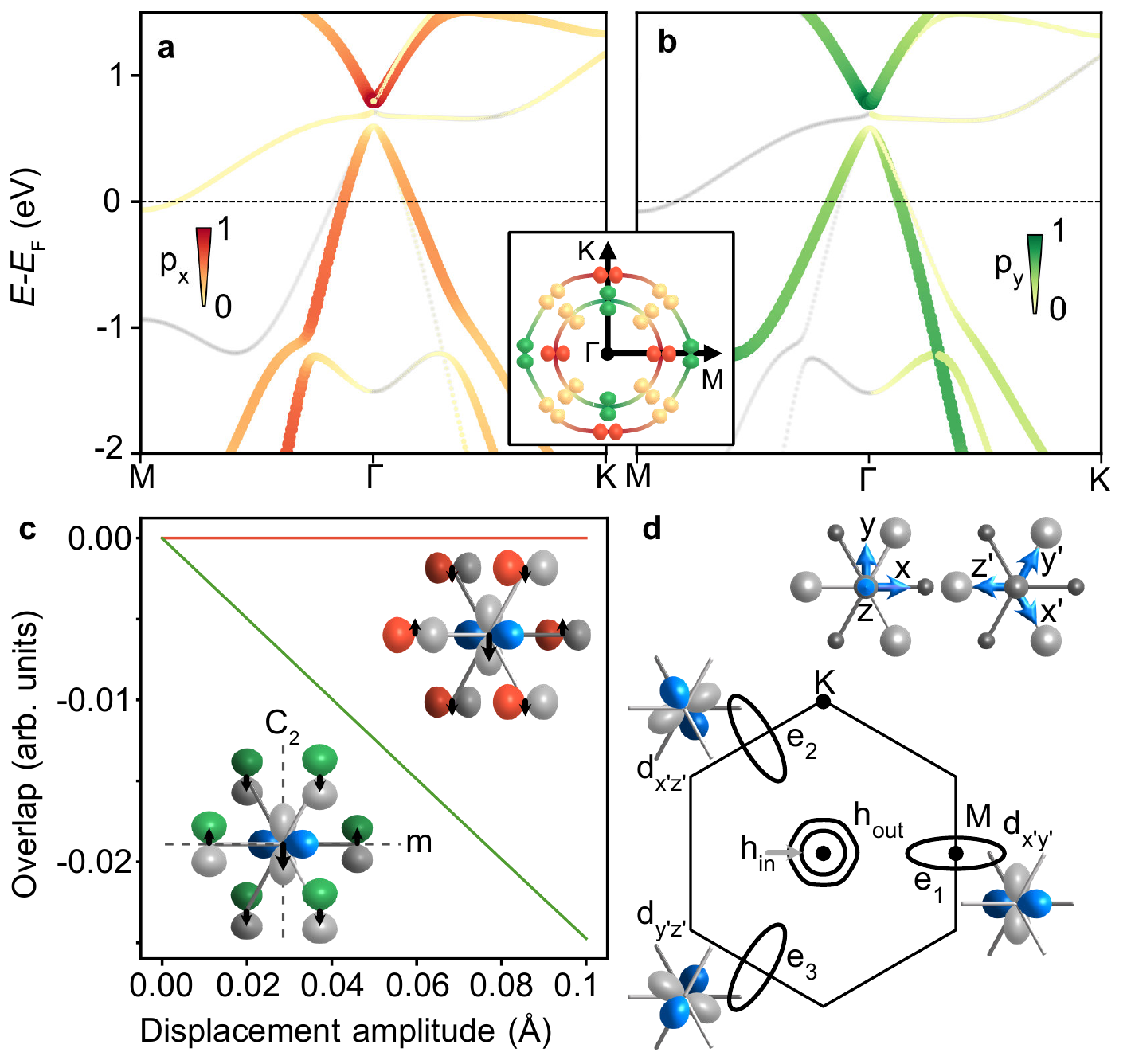}
	\caption{(a,b) Tight-binding band structure of ML-TiTe$_2$ neglecting SOC, revealing the $p_x$ (a, red colouring) and $p_y$ (b, green colouring) character along the M-$\Gamma$-K path. The inset shows the evolution of the valence band orbital character around the Fermi surface, indicating the radial and tangential orbital textures of the inner and outer bands, respectively. (c) Overlap integrals $\left\langle{}p_x|d_{x'y'}\right\rangle$ (red) and $\left\langle{}p_y|d_{x'y'}\right\rangle$ (green) as a function of the atom displacement caused by softening of an A$_u$ phonon mode at M (see insets; only the radial part of the wavefunctions are shown). (d) Schematic illustration of the two relevant coordinate systems, $xyz$ and $x'y'z'$, used to describe the orbital character in the tight-binding calculation within the crystal and octahedral basis, respectively. Below, a schematic of the 5 bands used in the minimal model is shown.}
	\label{fig4}
\end{figure}

\subsection*{Orbital-selective band hybridisation}
The second-derivative analysis also suggests the opening of spectral gaps in the reconstructed electronic structure. To investigate this in detail, to understand better how the electronic structure around the Fermi level evolves across the CDW transition, we performed high-resolution measurements at the $\Gamma$ point of the Brillouin zone. Here, the spectral weight is dominated by the valence bands (Fig.~\ref{fig3}(a)). Comparing the high- and low-temperature spectra, a small spectral gap can be seen to open in the vicinity of the Fermi level for the outer valence band, while no such gap is obvious for the inner valence band. To probe this quantitatively, we show the intensities of Lorentzian fits to MDCs of the outer and inner valence bands in Figs.~\ref{fig3}(b,c), respectively (see also Supplemental Fig.~S2 for details of the fitting). The intensities vary smoothly for the high-temperature measurements, with an overall decrease in intensity approaching the Fermi level reflecting the effects of transition matrix elements on the photoemission measurements, and also the influence of spectral broadening and rotational disorder of the grown films. This global intensity variation is still present at low temperature, but with an additional modulation leading to dips in fitted intensity at binding energies of $\approx50$~meV and $\approx100$~meV for the outer and inner bands, respectively. This is clearly visible in normalised difference plots of the fitted intensity at high and low temperatures (coloured arrows in Fig.~\ref{fig3}(d,e)). The suppressed spectral weight occurs at the energies at which the original valence and conduction bands intersect when backfolded in the low-temperature phase (inset of Fig.~\ref{fig3}(b)). This suggests their origin is due to a band hybridisation between the backfolded conduction and valence bands in the CDW state, opening band gaps that have remained hidden in the spectral broadening in measurements to date. Intriguingly, however, these gaps are highly asymmetric, with a pronounced minimum evident for the outer band, while only a small dip is observed at the band crossing energies for the inner valence band.

We will show below that this reflects an orbital- and symmetry-selective band hybridisation, providing key insight into the microscopic mechanisms of the stabilisation of CDW order in ML-TiTe$_2$, and its absence in the bulk. To demonstrate this, we introduce a minimal 5-band model capturing the low-energy electronic structure of ML-TiTe$_2$ in the CDW state (see Methods~\ref{ap:methods}):

\begin{equation}\label{ham}
\mathcal{H}=\begin{psmallmatrix}
 e_1          &       0           &          0          & \Delta c(\theta)                & \Delta s(\theta)\\ 
 0            &       e_2         &          0          & \Delta c(\theta-\frac{2\pi}{3}) & \Delta s(\theta-\frac{2\pi}{3})\\ 
 0            &       0           &          e_3        & \Delta c(\theta-\frac{4\pi}{3}) & \Delta s(\theta-\frac{4\pi}{3})\\ 
 \Delta c(\theta) & \Delta c(\theta-\frac{2\pi}{3}) & \Delta c(\theta-\frac{4\pi}{3}) &  h_{out}            & i\lambda_{SO} \\ 
 \Delta s(\theta) & \Delta s(\theta-\frac{2\pi}{3}) & \Delta s(\theta-\frac{4\pi}{3}) &  -i\lambda_{SO} & h_{in}
\end{psmallmatrix}
\end{equation}

where $e_i$ \{$i=1,2,3$\} represent the normal state dispersion of the elliptical electron pockets centered at neighbouring M points which are backfolded to $\Gamma$ in the low-temperature phase (Fig.~\ref{fig1}(g)), $h_i$ \{$i=out,in$\} are the two hole bands centered at $\Gamma$, spin-orbit coupling, $\lambda_{SO}$, is included for the Te-derived hole bands, and $c(x)$ and $s(x)$ represent $\cos(x)$ and $\sin(x)$, respectively.

\begin{figure*}
	\centering
	\includegraphics[width=\linewidth]{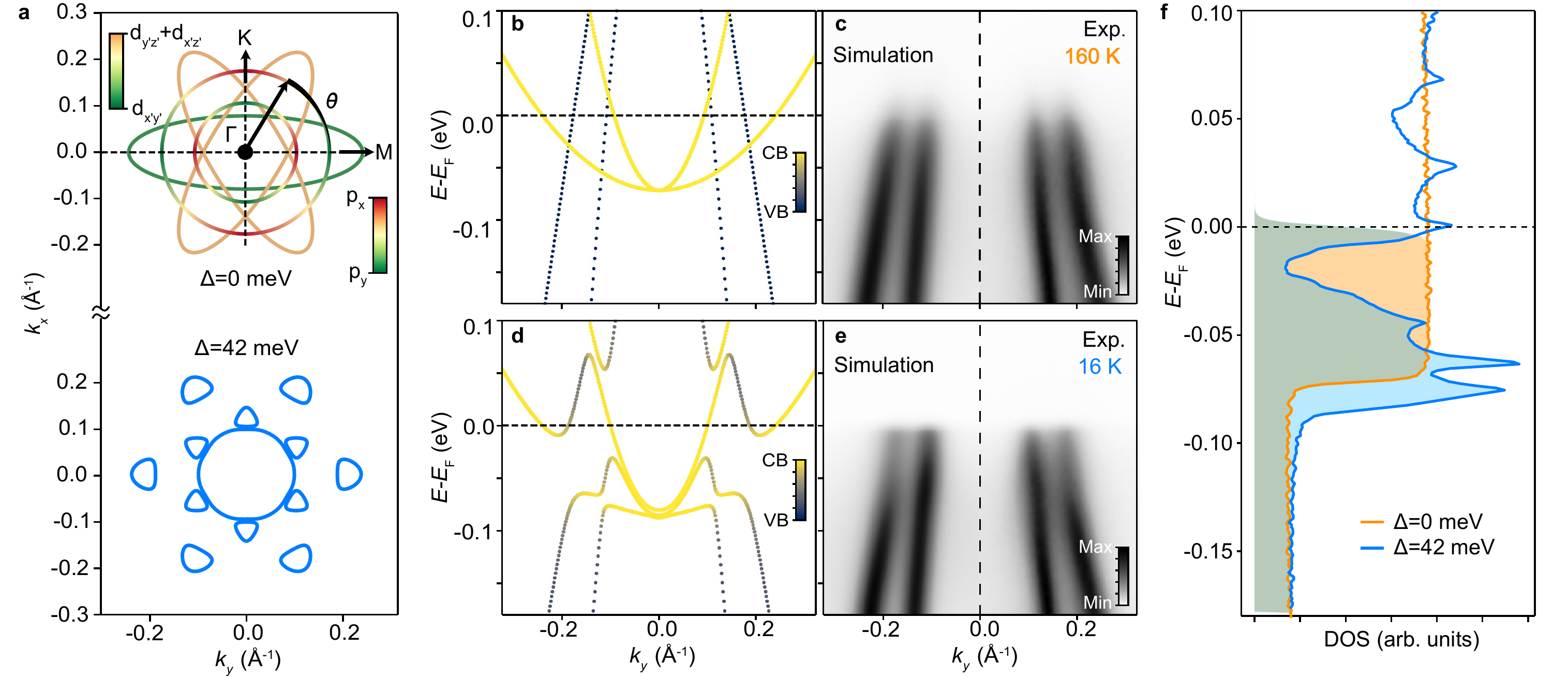}
	\caption{(a) Fermi surface calculated within our minimal model for ML-TiTe$_2$, with three electron pockets from M backfolded to $\Gamma$, but without hybridisation ($\Delta$ = 0~meV, top). As the hybridisation is turned on ($\Delta$ = 42~meV, bottom), the Fermi surface is partially gapped where allowed by symmetry. (b,d) Bare bands calculated from the minimal model along $\Gamma$-M for $\Delta$ = 0~meV and $\Delta$ = 42~meV, respectively, projected onto the character of the original conduction and valence states (blue-yellow colouring). (c,e) Comparison between the experimental data and simulated ARPES spectra from the model calculation in (b) and (d), respectively. (f) Calculated DOS from the model, indicating the effects of band hybridisation, providing a significant electronic energy gain in the occupied states (shaded region).}
	\label{fig5}
\end{figure*}

The key physics of the resulting band hybridisation between the backfolded conduction and valence states is included via an interaction term $\Delta$, whose strength is modulated by angle-dependent form factors, where $\theta$ is the angle of the momentum vector within the 2D plane, $\theta = \tan^{-1}(k_y/k_x)$. These form factors derive from the particular orbital textures of the valence states. Along $\Gamma$-M, our tight-binding calculations, informed from our DFT-calculated electronic structure and optimised to match the experimental normal-state dispersions (see Methods~\ref{ap:methods}), indicate that the inner valence band is derived mainly from Te 5$p_x$ orbitals, while the outer valence band has a dominant $p_y$ orbital character, where $x$ and $y$ are referenced to the global coordinate system (Fig.~\ref{fig4}(d)). Moving away from this direction, the orbital content rotates such that, for the perpendicular $\Gamma$-K direction, the inner band is predominantly of $p_y$ character, while the outer band is $p_x$-like. The inner and outer valence bands thus host radial and tangential orbital textures as shown in the inset of Fig.~\ref{fig4}(a,b), much like those recently uncovered in, {\it e.g.,} topological states of Bi$_2$Se$_3$
\cite{Cao2013}\cite{PhysRevLett.112.076802} and Rashba states of BiTeI \cite{Lewis2015}. Here, we will show that these orbital textures lead to a strong momentum-dependence of the allowed band hybridisation in the ordered state of TiTe$_2$.

To demonstrate this, we consider the overlap of the $p$ states with one of the three backfolded conduction band pockets, $e_1$, which has predominantly $d_{x'y'}$ symmetry in the octahedral basis (\{$x',y',z'$\}, Fig.~\ref{fig4}(d))~\cite{Kaneko2018PRB}. The situation for the other bands follows from the three-fold rotational symmetry of the system. To understand the hybridisation, we consider two core symmetries of the normal-state 1T crystal structure: a mirror plane, $m$, that is oriented along $\Gamma$-M, and a $C_2$ rotational symmetry axis oriented along $\Gamma$-K. The 3$d_{x'y'}$ orbital has even parity in both of these (see inset in Fig.~\ref{fig4}(c)), consistent with the $A_g$ irreducible representation of the electron pocket at M \cite{Huang2021}. In contrast, the state made from $p_y$ orbitals on the two chalcogen sites is even in $C_2$, but odd in $m$, while the opposite is true for the $p_x$-derived bands (see insets in Fig.~\ref{fig4}(c)). 

Considering at the $\Gamma$ point, where both of these symmetries are present for the high-temperature 1T crystal structure, the overlap integrals $\left\langle{p_x|d_{x'y'}}\right\rangle$ and $\left\langle{p_y|d_{x'y'}}\right\rangle$ are strictly zero, as a result of the opposite parity of these states under the $C_2$ rotation and the mirror symmetry, respectively. At the CDW transition, however, the softening of three $A_u$ phonon modes (Fig.~\ref{fig1}(h)) leads to a periodic lattice distortion~\cite{Kaneko2018PRB} which breaks the mirror symmetry, while preserving the $C_2$ rotational symmetry (the corresponding atomic displacements arising from the softening of one of the three phonon modes are indicated by the black arrows in the insets of Fig.~\ref{fig4}(c)). The hybridisation of the $p_x$- and $d$-derived states, $\left\langle{p_x|d_{x'y'}}\right\rangle$, remains zero due to the $C_2$ symmetry. However, the mirror-symmetry enforced constraint of a lack of hybridisation between the $p_y$- and $d$-derived states in the normal state is relaxed in the distorted structure. This is evident from our calculated overlap integrals in Fig.~\ref{fig4}(c), where the hybridisation matrix element between $p_y$ and $d_{x'y'}$ tesseral harmonics increases linearly with the amplitude of the atomic displacements within the periodic lattice distortion, while $\left\langle{p_x|d_{x'y'}}\right\rangle$ remains zero irrespective of the atomic displacement.

While these arguments are formally valid only at $\Gamma$ where both the symmetries are present, they provide the basis for an understanding of the band hybridisation throughout the Brillouin zone, where the $C_2$ (in both the normal and low-temperature states) and $m$ (in the normal state) symmetries are present along the entire $\Gamma$-K and $\Gamma$-M directions, respectively. In particular, they indicate that new channels for band hybridisation between the backfolded valence and conduction band pockets are only opened up by the periodic lattice distortion where conduction bands of $d_{x'y'}$ character intersect valence bands of $p_y$-orbital character (and symmetry-equivalent versions of these for the other conduction band pockets).

Combined with the momentum-dependent orbital textures of the valence bands discussed above, this significantly constrains and modulates the hybridisation in the 2D $k$-space, as captured in the minimal model of Equation~\ref{ham}, and shown from our model calculations in Fig.~\ref{fig5}(a,b,d). While the strict symmetry protections outlined above are weakened by spin-orbit coupling on the Te site, which mixes the $p_x$ and $p_y$ states, it is clear how the above symmetry constraints lead to a strong asymmetry in the band gaps which open up between crossings of, {\it e.g.}, the $e_1$ and $h_{in}$ and $h_{out}$ bands for finite values of hybridisation strength $\Delta$ (Fig.~\ref{fig5}(b,d)). The overall hybridisation scheme derived from our model is qualitatively consistent with the results of recent DFT calculation of ML-TiTe$_2$ in the $(2\times2)$ phase~\cite{Sky_Zhou_2020,Guster_2018}: although the energy gap is overestimated in such calculations, they predict a weakly hybridised band laying in between the CDW gap as in our model.

To validate our model and benchmark it against our experimental data, we simulate the resulting ARPES spectra expected from these hybridised bands (as described in Methods~\ref{ap:methods}). We find a good agreement between our simulated and measured spectra (Fig.~\ref{fig5}(c,e)) taking $\Delta = 42\pm10$~meV. This is a significantly weaker interaction strength than for the sister compound ML-TiSe$_2$, where $\Delta\approx100$~meV can be directly estimated from the experimental data~\cite{Watson_2020}, entirely consistent with the weaker nature of the instability predicted by our calculations of the lattice dynamics discussed above (Fig.~\ref{fig1}(i)) and with the lower $T_c$. Crucially, our simulated spectra well reproduce the suppression of spectral weight in the ordered phase identified in fits to MDCs of our measured spectra (see green lines in Fig.~\ref{fig3}(d,f)), including the pronounced asymmetry in spectral weight suppression at the crossing of $e_1$ and $h_{in}$ and $h_{out}$. This, therefore, provides direct experimental evidence for the symmetry and orbital-selectivity of the band hybridisation at the CDW phase transition in ML-TiTe$_2$.

\begin{figure}[!b]
	\includegraphics[width=\columnwidth]{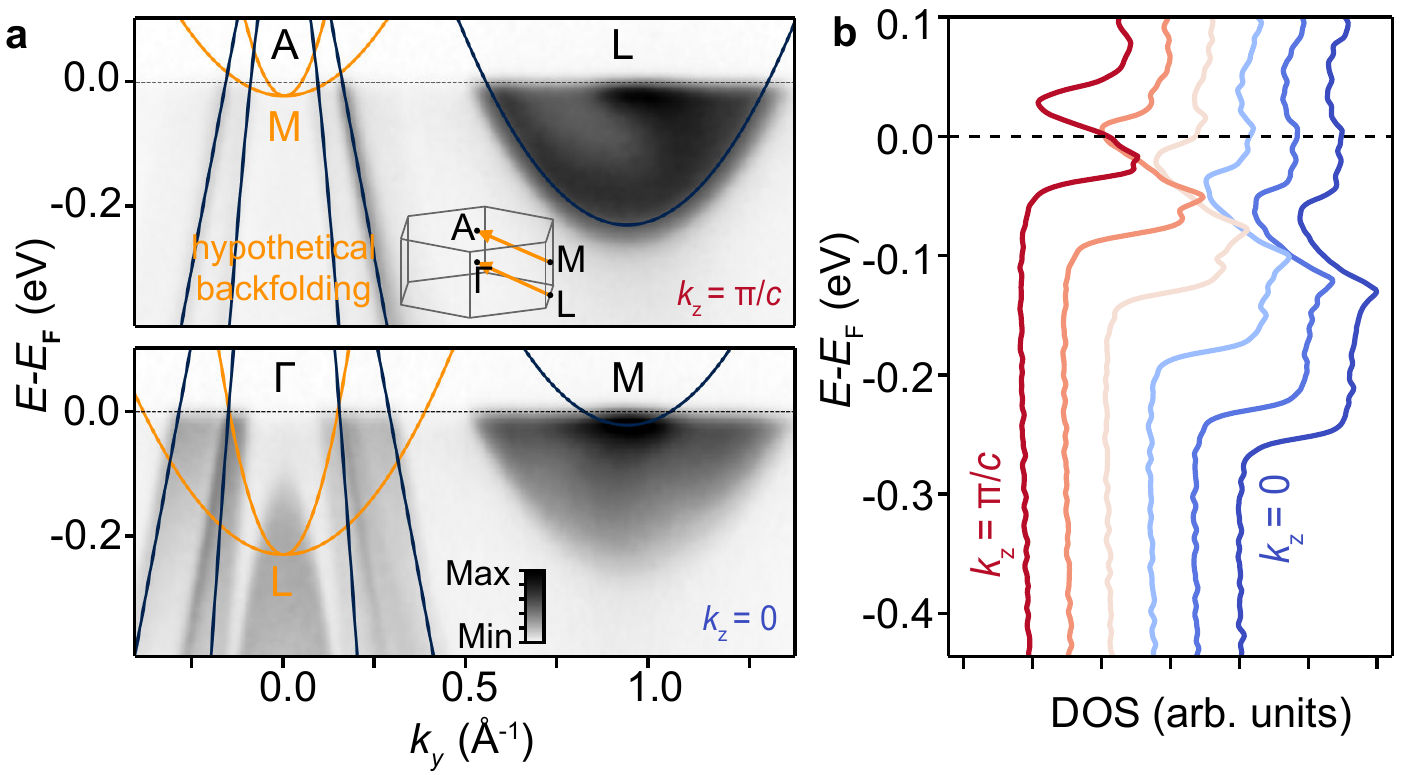}
	\caption{(a) ARPES measurements ($T=14$~K) of bulk TiTe$_2$ at $k_z$ = $\pi/c$ ($h\nu=144$~eV, top) and at $k_z$ = 0 ($h\nu=121$~eV, bottom). The orange lines show the backfolded bands that would be expected from a hypothetical $(2\times2\times2)$ periodic lattice distortion as is observed for the sister compound TiSe$_2$. No such CDW backfolding is observed in the experiment. (b) Calculated density of states (DOS) from a model considering such a hypothetical $(2\times2\times2)$ backfolding, and assuming a hybridisation strength equal to the one estimated for the monolayer. The significant size mismatch of the backfolded electron and hole bands, evident in (a), leads to a DOS suppression from the band hybridisation which disperses from below to above the Fermi level as a function of $k_z$, suppressing the associated electronic energy gain.}
	\label{fig6}
\end{figure}

\subsection*{CDW energetics: from monolayer to bulk}
Having validated the essential properties of our model, we can use this to provide new insight on the key question of why the CDW phase becomes stable in ML-TiTe$_2$, while it is absent for the bulk. First, we note that there is in fact a significant electronic energy gain that results from the above band hybridisations at the CDW transition in the monolayer system. While the hybridisation gaps along $\Gamma$-M occur largely below the Fermi level, at other angles around the Fermi surface, the gaps open at the Fermi level itself (Fig.~\ref{fig5}(a)) partially gapping the Fermi surface. Indeed, calculating the density of states for the normal and hybridised states (Fig.~\ref{fig5}(f)) it is evident how the hybridisation in ML-TiTe$_2$ significantly lowers the total electronic energy over an extended bandwidth comparable to the band overlap in the normal state. We thus conclude that this electronic energy gain is ultimately the driver of the CDW transition in ML-TiTe$_2$. 

Given this, one might assume that a similar band hybridisation in the bulk would stabilise a CDW there, where none is known to exist experimentally, nor is one found as an instability in our DFT calculations (Fig.~\ref{fig1}(d)). However, there are two key distinctions in bulk TiTe$_2$ as compared to the ML case. First is a sizeable out-of-plane dispersion of the electronic states, for both the valence and conduction bands, as shown in Fig.~\ref{fig6}(a). The conduction band states yield large Fermi pockets for $k_z=\pi/c$, while they are barely occupied for $k_z=0$. Though the valence state dispersion is weaker, the Fermi wavevectors nonetheless exhibit a notable decrease with increasing $k_z$. For a $(2\times2)$ ordering as in the monolayer case, the backfolded electron- and hole-like Fermi pockets are completely mismatched in size, and thus unlikely to lead to any significant electronic energy gain as for the monolayer. Even for a $(2\times2\times2)$ instability as in the sister compound TiSe$_2$~\cite{DiSalvo1976PRB}, the pockets are still significantly mismatched (Fig.~\ref{fig6}(a)), causing the energies at which the hypothetical backfolded bands overlap to disperse strongly as a function of the out-of-plane momentum. Including a hybridisation between these bands in the same manner as we did for the monolayer case again leads to characteristic dips in the DOS, but these DOS suppressions now move from the occupied to the unoccupied states as a function of the out-of-plane momentum, $k_z$ (Fig.~\ref{fig6}(b)). As a result, this significantly reduces the effectiveness of the band hybridisation here to lower the total electronic energy of the system as compared to the normal state.

\begin{figure*}
    \includegraphics[width=\linewidth]{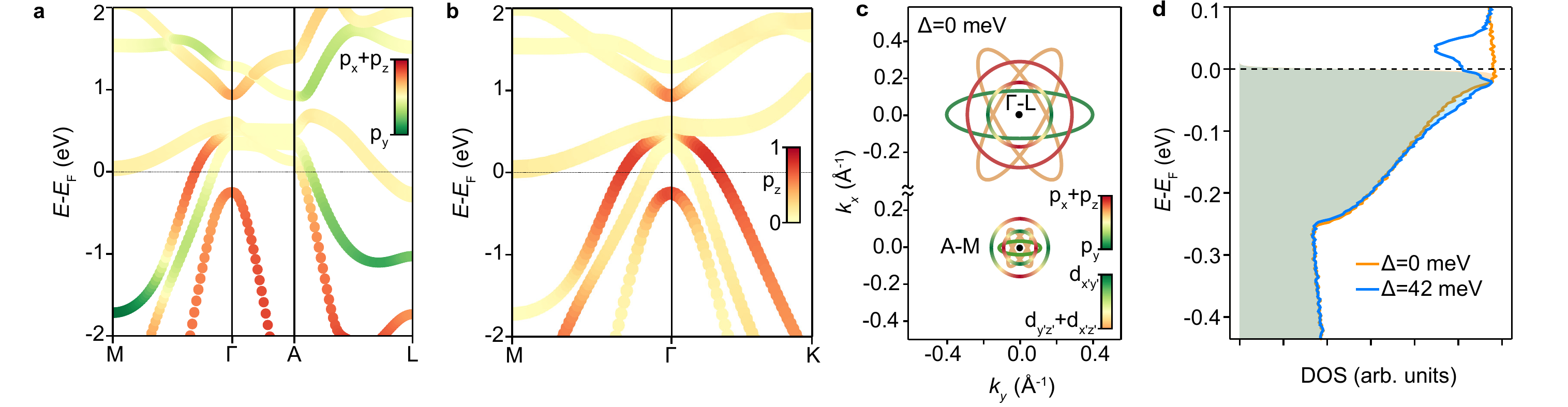}
	\caption{(a,b) DFT band structure projected onto the atomic $p$ orbitals (red-green colouring) showing (a) an inversion of the orbital character of the valence bands for the $\Gamma$-plane as compared to the A-plane. (b) Isotropic $p_z$ character at the Fermi level of the outer valence band in the $\Gamma$-plane. (c) Fermi surface of bulk TiTe$_2$ from our minimal model, sliced in the $k_x-k_y$ plane and including a hypothetical $(2\times2\times2)$ backfolding ({\it i.e.} from $\Gamma$ to L and A to M). The colors indicate the orbital character of the Ti 3$d$-derived and Te 5$p$-derived electron and hole pockets, respectively. (d) DOS of bulk TiTe$_2$ assuming such a backfolding and symmetry-allowed band hybridisations of strength $\Delta$ = 0~meV and $\Delta$ = 42~meV, indicating negligible electronic energy gain for the CDW state in the bulk.}
	\label{fig7}
\end{figure*}

Moreover, our DFT calculations indicate that the significant out-of-plane dispersion in fact leads to a $k_z$-dependent band inversion of the $d$ and $p$-derived states along the $\Gamma$-A direction, of the form known from other families of TMDs where it generates bulk Dirac points and topological surface states~\cite{Bahramy2018}. As shown in Fig.~\ref{fig7}(a), this causes the orbital character of the inner and outer valence bands to become inverted between the $\Gamma$- and the A-plane, switching approximately mid-way through the bulk Brillouin zone along $k_z$. Additionally, $p_z$ character is also mixed with the $p_x$ states. The $p_z$ orbitals exhibit odd parity with respect to the $C_2$ axis, forbidding their hybridisation with the electron pockets as for the $p_x$-derived states. The in-plane rotational symmetry of the $p_z$ orbitals is also reflected via an isotropic $p_z$-dominated character of the outer Fermi surface sheet for $k_z$ = 0, shown in Fig.~\ref{fig7}(b,c). This suppresses the hybridisation with the electron pockets even away from the $\Gamma$-M direction. We have incorporated all of these effects ($k_z$-induced size mismatch, switch of orbital selectivity mediated by the band inversion, and the additional incorporation of the $p_z$ state) into our minimal model (see Methods~\ref{ap:methods}). We find that only minimal changes in the occupied DOS then occur between the high-temperature state and when considering a $(2\times2\times2)$ structure with the same hybridisation strength that we found for ML-TiTe$_2$ (Fig.~\ref{fig7}(d)). Our calculations thus demonstrate that the electronic driving force for the CDW instability present in the monolayer is effectively completely removed in the bulk case.

\section*{Discussion}
Our work thus points to a key role of orbital-selective band hybridisation in dictating whether CDW instabilities can occur in TiTe$_2$. Understanding microscopic details of electron-phonon coupling will undoubtedly be important to gain a full understanding of the CDW ordering. Nonetheless, our experimental measurements and model calculations indicate how significant electronic energy gains are derived from band hybridisation arising from a $(2\times2)$ lattice distortion in the monolayer case, which are almost completely suppressed due to the three-dimensionality, and associated band inversions, in the bulk case. This provides a natural picture to explain the emergence of CDW order in monolayer TiTe$_2$ as well as its intriguing thickness-dependence, and may bring new insight to understand a recently-observed enhanced CDW order in Moir{\'e} superlattices of TiTe$_2$-based heterostructures~\cite{zhao2022}. More generally, the model developed here will provide a natural framework through which to interpret and understand the low-energy electronic structure evolution in the sister compounds TiSe$_2$ and ZrTe$_2$, which host the same crystal structure and similar electronic structures, and where the nature of their CDW-like order is under intense current debate~\cite{Rossnagel2002,kogar,song2022}.

\begin{figure*}
	\centering
	\includegraphics[width=\linewidth]{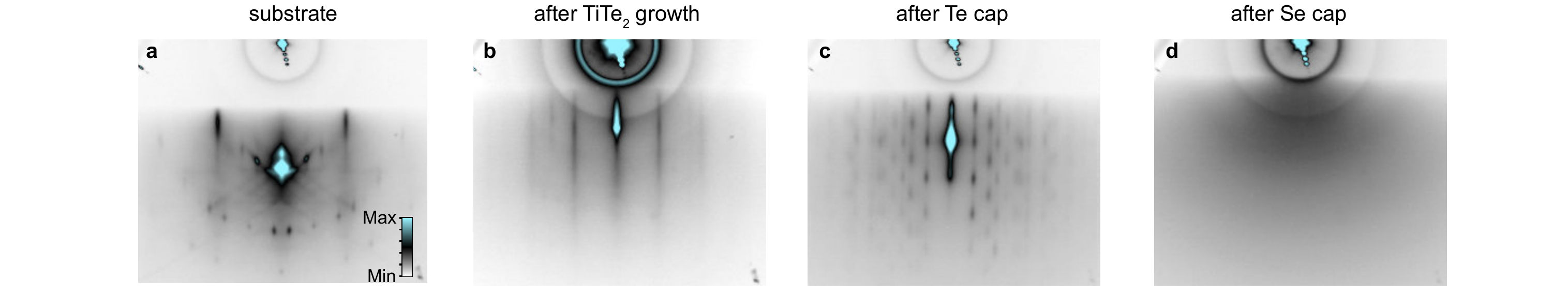}
	\caption{RHEED images at different steps of the MBE growth of ML-TiTe$_2$ and capping.}
	\label{RHEED}
\end{figure*}

\section*{Methods}\label{ap:methods}
\textbf{Sample preparation:}
ML-TiTe$_2$ was grown by molecular beam epitaxy (MBE) on bilayer graphene-terminated SiC wafers using the method described in \cite{Rajan2020PRM}. The bilayer graphene termination of the SiC wafer was achieved by direct current heating of SiC wafers at $1500^\circ$C in a dedicated high-vacuum chamber equipped with a pyrometer to check the temperature. These substrates were degassed at $600^\circ$C before the growth was commenced. The epitaxy of ML-TiTe$_2$ was performed in a highly Te-rich environment (Ti:Te flux ratio $\sim\!10^{3}$) at a substrate temperature of $400^\circ$C as measured by a thermocouple placed behind the substrate. The growth was stopped after 70 minutes while the graphene RHEED streaks are still visible (see Fig.~\ref{RHEED}(a,b)) and before the onset of bilayer formation, as confirmed from the lack of any bilayer splitting in our subsequent ARPES measurements of the electronic structure. The sample was then capped with Te for 20 min resulting in a spotty RHEED pattern (Fig.~\ref{RHEED}(c)), indicating a crystalline Te layer completely covering the underlying TiTe$_2$ mono-layer. An additional amorphous Se capping was deposited to protect the Te layer against oxidation in air, and was deposited until the Te signal disappeared completely (Fig.~\ref{RHEED}(d)).
After growth, the film was capped with an initial Te layer and a subsequent amorphous Se layer which were removed by {\it in situ} annealing in the CASSIOPEE endstation at SOLEIL synchrotron immediately prior to performing ARPES measurements. Additional measurements on bulk TiTe$_2$ single crystals were performed on freshly-cleaved single-crystal samples.
\\

\textbf{DFT:}
DFT calculations were performed using the Vienna Ab Initio Simulation Package (VASP)~\cite{Kresse1993,Kresse1996a,Kresse1996b} in order to investigate the lattice dynamical properties of TiSe$_2$ and TiTe$_2$ (bulk and monolayer). The interactions between the core and valence electrons were described using the projector augmented wave (PAW) method~\cite{PhysRevB.50.17953}, with Ti 3$p^6$4$s^2$3$d^2$, Se 4$s^2$4$p^4$ and Te 5$s^2$5$p^4$ treated as valence electrons. The HSE06 (Heyd-Scuzeria-Ernzerhof)~\cite{Krukau,Heyd} screened hybrid functional was employed to describe the exchange and correlation functional. Within HSE06, the exchange interaction is split into short-range (SR) and long-range (LR) parts, with 25$\%$ of the SR exchange modelled using the exact non-local Fock exchange and the remainder of the contributions coming from the PBE (Perdew-Burke-Ernzerhof)~\cite{PhysRevLett.77.3865} functional. The hybrid functional calculations are computationally very demanding, but are expected to provide a better description of structure and phonon frequencies~\cite{PhysRevLett.108.259601}. The empirical correction scheme of Grimme’s DFT-D3 method~\cite{Grimme} was also employed along with the hybrid functional to model the van der Waals interactions between the layers, as it has been successful in accurately describing the geometries of several layered materials in the past~\cite{Moellmann_2012,PARK2015885,PhysRevMaterials.2.034005}. For the monolayers, a vacuum layer of 12~\AA{} was included to avoid interaction between the periodic images. Geometry optimisations were performed by setting the plane-wave energy cutoff at 350 eV for each structure, alongside $\Gamma$ centred $k$-point meshes of $16\times16\times10$ (TiSe$_2$-bulk), $16\times16\times8$ (TiTe$_2$-bulk), and $16\times16\times1$ for the monolayers. A full relaxation of atomic positions, cell volume and shape was performed for the bulk structures until the residual forces acting on all the atoms were found to be less than 10$^{-4}$ eV/Å. The same convergence criterion for forces was used for relaxing the internal atomic coordinates of the monolayers. The self-consistent solution to the Kohn-Sham equations was obtained when the energy changed by less than 10$^{-8}$ eV. The phonon calculations were performed using the finite-displacement method (FDM)~\cite{PhysRevLett.78.4063,Kresse_1995} as implemented in the Phonopy package~\cite{PhysRevB.78.134106,phonopy}. The $k$-point sampling meshes were scaled accordingly as appropriate for the supercell calculations. We also mapped the potential energy curves spanned by the imaginary frequency modes in the undistorted structures to estimate the barrier associated with CDW transition. For this purpose, we used the ModeMap code to create a series of structures with atomic displacements along the mode eigenvectors over a range of amplitudes “frozen in”. This was followed by performing single-point energy calculations on the modulated structures to obtain the double-well potential spanned by each mode.\\

\textbf{Tight binding:}
The band structure of ML-TiTe$_2$ was calculated adapting the Slater-Koster tight-binding model used for TiSe$_2$ in Ref.~\cite{Kaneko2018PRB}. Accordingly, we defined the orbital basis in the octahedral coordinates \{$x'y'z'$\} in Fig.~\ref{fig4}(d), considering the three Te 5$p$ orbitals $p$~\{$p_x', p_y', p_z'$\} for each of the two chalcogen sites (Te(1) and Te(2)), and the five Ti 3$d$ orbitals $d\gamma$~\{($d_{x'^2−y'^2} , d_{3z'^2−r'^2} )$\} and $d\epsilon$~\{($d_{x'y'} , d_{y'z'}, d_{x'z'})$\}. The hopping parameters and on-site energies reported in Table \ref{table1} were determined by initially fitting the tight binding model along M-$\Gamma$-K to the DFT band structure, and subsequently refining the fits using our experimental ARPES data of the undistorted phase (see Supplemental Fig.~S3). An on-site spin–orbit coupling term on the Te sites is included ($\lambda$ = 0.38 eV), optimised from fitting the dispersion on the high temperature ARPES spectrum. \\
\begin{table}
\caption{Tight binding parameters for ML-TiTe${_2}$}
    \def\arraystretch{1.8}
    \setlength\tabcolsep{10pt}
    \begin{tabular}{@{}lll@{}}
    \toprule
    \multicolumn{3}{c}{On-site energies (eV)}  \\
    \hline
    $e(p)$ = -1.958  &  $e(d\epsilon)$ = 0.256 &  $e(d\gamma)$ = 0.721\\
    \toprule
    \multicolumn{3}{c}{Hopping parameters (eV)}  \\ 
    \hline

    $t(pp\sigma)_1$ =  0.904   & $t(pp\sigma)_2$ = 0.581   & $t(pd\sigma)$ = -1.618 \\
    $t(pp\pi)_1$ = -0.139      & $t(pp\pi)_2$ = 0.075      & $t(pd\pi)$ = 0.992 \\
    $t(dd\sigma)$ = -0.427     & $t(dd\pi)$ = 0.216        & $t(dd\delta)$ =  -0.070 \\

    \toprule
    \end{tabular}
\label{table1}
\end{table}

\textbf{Overlap evaluation:}
The orbital overlaps in Fig.~\ref{fig4}(c) were calculated considering the tesseral harmonics of the six Te $p$ orbitals at the vertex of the TiTe$_6$ octahedron defined in the global coordinate system, $\{x,y,z\}$ and the Ti $d_{x'y'}$ orbital at the centre of the TiTe$_6$ octahedron defined in the octahedral coordinate system $\{x'y'z'\}$. The overlap integral was evaluated numerically on a $(121\times121\times121)$ cubic mesh. We simulate the effect of the phonon mode softening by displacing the atoms along the direction indicated by the black arrows in Fig.~\ref{fig4}(c), following the distortion pattern reported for TiSe$_2$ in Ref.~\cite{DiSalvo1976PRB}. Since Ti is lighter that Te, we assume the Ti displacement is 9 times larger that the Te ones.
\\

\textbf{Minimal model:}
As basis for the Hamiltonian in Equation~\ref{ham} we choose the five bands involved in the hybridisation near the Fermi level. The three electron pockets ($e_i$, $i ={1,2,3}$) are parametrised as 2D elliptic paraboloids, while the two hole bands as circular non-parabolic bands:
\begin{equation}
e_1 = e_c+\mu k_x^2+\nu k_y^2
\end{equation}
\begin{equation}
e_3 = e_c+\mu(-\frac{1}{2}k_x+\frac{\sqrt{3}}{2}k_y)^2+\nu(-\frac{\sqrt{3}}{2}k_x-\frac{1}{2}k_y)^2
\end{equation}
\begin{equation}
e_3 = e_c+\mu(-\frac{1}{2}k_x-\frac{\sqrt{3}}{2}k_y)^2+\nu(\frac{\sqrt{3}}{2}k_x-\frac{1}{2}k_y)^2
\end{equation}
\begin{equation}
h_{in} = e_h+(\sqrt{2}-\sqrt{2+4 \alpha_{in} \mu{}k^2})/(2 \eta_1)
\end{equation}
\begin{equation}
h_{out} = e_h+(\sqrt{2}-\sqrt{2+4 \alpha_{out} \mu{}k^2})/(2 \eta_2)
\end{equation}
where $e_h$ and $e_c$ are the band minimum energy for the valence and conduction bands respectively; $\mu$ and $\nu$ are the effective masses along the major and minor axis of the elliptical parabola, and $\alpha_{in}$ and $\alpha_{out}$ are the non-parabolic terms for the inner and outer valence band following the Kane model~\cite{yu2005fundamentals}. All these coefficients were determined by fitting the simulated spectrum with $\Delta = 0$ on the experimental data at $T= 160$~K shown in Fig.~ \ref{fig5}(c).

In order to extend the model to the bulk case we introduced and additional cosine dispersion in k$_z$ for the onsite energies:
\begin{equation}
e_c = [e_c(L)-e_c(M)]cos(k_z)+[e_c(L)+e_c(M)]/2
\end{equation}
\begin{equation}
e_h = [e_h(\Gamma)-e_h(A)]cos(k_z)+[e_h(\Gamma)+e_h(A)]/2
\end{equation}
where the values of $e_c$ at M and L and $e_h$ at $\Gamma$ and A were determined by fitting the ARPES data in Fig~\ref{fig6}(a). To ensure charge neutrality, a shift of the chemical potential across the CDW transition was taken into account, simulating how the occupied DOS depends on the hybridisation strength ($\Delta$) at constant $T= 16$~K (see Supplemental Fig.~S4 for details).
\\

\textbf{ARPES simulation:}
The ARPES simulations in Fig.~\ref{fig5}(c,e) were performed taking into account the intensities of the 5 bands ($e_1, e_2, e_3, h_{in}, h_{out}$) considered in the minimal model plus a background with third-order polynomial along the momentum $k$ and a linear dispersion along the energy axis $\omega$. Each band intensity $I_B(k,\omega)$ takes the form of:
\begin{equation}
I_B(k,\omega) = [M(k)
A(k,\omega) \textit{f}(T,\omega)]\ast R(\Delta k,\Delta \omega)
\end{equation}
where $M$ is the matrix element, $A$ is the spectral function, $\textit{f}(T,\omega)$ is the Fermi Dirac distribution at temperature $T$. The entire expression is then convolved with a 2D Gaussian ($R$) to simulate the experimental energy and momentum resolution. $M$ depends on the band character $C_b$ and can be approximated to have a linear dependence on $k$: 
\begin{equation}
M = \sum_{b=1}^{5} m_bC_b^2k + n_bC_b^2 .
\end{equation}
$A$ takes into account the Lorentzian broadening of the band intensity due to impurity scattering: 
\begin{equation}
A(k,\omega) = \frac{\Sigma_b}{(\omega-e_b(k)^2+\Sigma_b^2}
\end{equation}
where $e_b$ is the bare band dispersion calculated by the minimal model and $\Sigma_b$ is the Lorentzian FWHM.
For measurements performed at the Brillouin zone centre, the matrix element can be expected to approximately follow that of original valence band character~\cite{Grioni2000,SunkoSciAdv} (dark blue in Fig.~\ref{fig5}(b,d)). We note that for both the high and low $T$ spectra the band intensities decrease monotonically form -0.16 eV up to the Fermi level. Since it is not a temperature-dependent feature we do not consider it related to the CDW transition, while it might be due to the presence of defects in the ML-TiTe$_2$ or azimuthal disorder in the grown film. In order to have a better agreement with the experimental data we include a temperature-independent intensity decay, $I_{dec}$, of the from:
\begin{equation}
  I_{dec}=\begin{cases}
    cos\left[\frac{\pi(\omega - E_i)}{2(E_f-E_i)}\right], & \text{if $\omega \leq E_i$}\\
    1, & \text{if $\omega>E_i$}
  \end{cases}
\end{equation}
where the coefficients $E_f$ and $E_i$ were determined fitting the high temperature data and kept constant for all the simulations at different temperatures. Thus, this contribution is cancelled out when we calculate the intensity difference between the high and low temperature data as in Fig.~\ref{fig3}.
\\

\section*{Acknowledgments}
We thank Sebastian Buchberger, Brendan Edwards, Lewis Hart, Chris Hooley, Federico Mazzola,  Martin McClaren, Philip Murgatroyd and Gesa Siemann for useful discussions and technical assistance. We gratefully acknowledge support from the Leverhulme Trust and the Royal Society. Via membership of the UK's HEC Materials Chemistry Consortium, which is funded by the EPSRC (EP/L000202, EP/R029431, EP/T022213), this work used the ARCHER2 UK National Supercomputing Service (www.archer2.ac.uk) and the UK Materials and Molecular Modelling (MMM) Hub (Thomas – EP/P020194 \& Young – EP/T022213). W.R. is grateful to University College London for awarding a Graduate Research Scholarship and an Overseas Research Scholarship. O.J.C. and K.U. acknowledge PhD studentship support from the UK Engineering and Physical Sciences Research Council (EPSRC, Grant Nos.~EP/K503162/1 and EP/L015110/1). I.M. and E.A.-M. acknowledge studentship support from the International Max-Planck Research School for Chemistry and Physics of Quantum Materials. S.R.K. acknowledges the EPSRC Centre for Doctoral Training in the Advanced Characterisation of Materials (CDT-ACM, EP/S023259/1) for funding a PhD studentship. The MBE growth facility was funded through an EPSRC strategic equipment grant: EP/M023958/1. We thank SOLEIL synchrotron for access to the CASSIOPEE beamline (proposal Nos. 20181599 and 20171202). The research leading to this result has been supported by the project CALIPSOplus under Grant Agreement 730872 from the EU Framework Programme for Research and Innovation HORIZON 2020.

\section*{Contributions}
M.D.W., A.R., O.J.C., K.U., I.M., E.A.M. and P.D.C.K. measured the ARPES data, which was analysed by T.A.\\
W. R. and S. R. K. performed the DFT calculations, which were analysed by W.R., T.A., D.O.S and P.D.C.K.\\
T.A. performed the tight-binding calculations, ARPES simulations and matrix element calculations.
M.W., T.A. and P.D.C.K. developed the minimal model.
A.R. and A.D. grew the monolayer samples. K.R. grew the single crystal samples. P.L.F. and F.B. maintained the CASSIOPEE beam line and provided experimental support. P.D.C.K. conceived and led the project. T.A. and P.D.C.K. wrote the manuscript with contributions from all authors.

\section*{Competing interests}
The authors declare no competing interests.


%

\end{document}